%% file: main.tex
\documentclass[conference]{IEEEtran}

\usepackage[noadjust]{cite}
\usepackage{multirow} 
\usepackage{algpseudocode}
\usepackage{algorithm}
\usepackage{rotating}
\usepackage{kantlipsum} 
\usepackage{commath}
\allowdisplaybreaks
\usepackage{mathtools}  
\usepackage{verbatim}
\usepackage{xr-hyper} 
\usepackage{enumitem}

\usepackage{epstopdf}
\usepackage{wrapfig}
\usepackage{latexsym}
\usepackage{amssymb}
\usepackage{amsthm}
\usepackage{amsfonts}
\usepackage{amsmath} 
\usepackage{graphicx}
\usepackage{latexsym}
\usepackage{booktabs}
\usepackage{support-caption} 
\usepackage{subcaption} 
\usepackage{xspace}

\usepackage[normalem]{ulem} 

\usepackage[T1]{fontenc}

\usepackage{hyperref}

\ifCLASSINFOpdf
\else
\fi
\newcommand{\dq}[1]{``#1''}

\usepackage[dvipsnames]{xcolor}

\newif\ifcommentson
\commentsontrue

\newif\ifextended
\newif\ifshortver

\extendedtrue

\newcommand{\extended}[1]{\ifextended \ifshortver \textcolor{purple}{#1} \else \textcolor{black}{#1} \fi  \fi}
\newcommand{\shortver}[1]{\ifshortver \ifextended \textcolor{blue}{#1} \else \textcolor{black}{#1} \fi \fi}

\newcommand{\parbreak}{\newline\hspace*{\parindent}}

\newif\ifrevision

\begin{document}

\bstctlcite{IEEEexample:BSTcontrol}


\title{Green Distributed AI Training:\\Orchestrating Compute Across\\Renewable-Powered Micro Datacenters}

\author{
\IEEEauthorblockN{
Giuseppe Tomei\IEEEauthorrefmark{3},
Andrea Mayer\IEEEauthorrefmark{1}\IEEEauthorrefmark{4},
Giuseppe Alcini\IEEEauthorrefmark{2},
Stefano Salsano\IEEEauthorrefmark{1}\IEEEauthorrefmark{2}
}
\IEEEauthorblockA{
\IEEEauthorrefmark{1}University of Rome Tor Vergata, Italy,
\IEEEauthorrefmark{2}CNIT, Italy,
\IEEEauthorrefmark{4}COMMON NET, Italy,
\IEEEauthorrefmark{3}proxim AI, Italy\\
Email: giuseppe@proxim.ai, andrea@common-net.org, giuseppe.alcini@cnit.it, stefano.salsano@uniroma2.it
}
\extended{\textbf{Extended version of the accepted paper - v02 - March 2026}}
}

\markboth{Journal of \LaTeX\ Class Files,~Vol.~14, No.~8, August~2015}%
{Shell \MakeLowercase{\textit{et al.}}: Bare Demo of IEEEtran.cls for IEEE Journals}
%



\maketitle



%
\IEEEpeerreviewmaketitle

\input{main-core-camera-ready}

\end{document}

%% file: main-core-camera-ready.tex
\shortver{
\begin{abstract}
The growing demand for AI workloads is driving unprecedented energy consumption, exacerbated by curtailed renewable energy and the poor energy proportionality of large datacenter nodes for modest-scale jobs. We propose a distributed network of renewable-powered micro-datacenters that (i) achieve lower system-level Joules-per-FLOP for a wide class of single-GPU workloads and (ii) are co-located with renewable sources to exploit zero-carbon, curtailed energy. This paradigm depends critically on the ability to migrate live AI workloads to follow intermittent energy availability. The main contribution of this work is a quantitative \emph{Feasibility Domain Analysis} that defines when such migration is viable under time and energy constraints. Within these boundaries, we design and evaluate a feasibility-aware orchestration framework that enables effective distributed AI training across renewable-powered micro-datacenters.
\end{abstract}
}

\extended{
\begin{abstract}
The accelerating expansion of AI workloads is colliding with an energy landscape increasingly dominated by intermittent renewable generation. While vast quantities of zero-carbon energy are routinely curtailed, today's centralized datacenter architectures remain poorly matched to this reality in both energy proportionality and geographic flexibility. This work envisions a shift toward a distributed fabric of renewable-powered micro-datacenters that dynamically follow the availability of surplus green energy through live workload migration.
\parbreak
At the core of this vision lies a formal feasibility-domain model that delineates when migratory AI computation is practically achievable. By explicitly linking checkpoint size, wide-area bandwidth, and renewable-window duration, the model reveals that migration is almost always energetically justified, and that time-not energy-is the dominant constraint shaping feasibility. This insight enables the design of a feasibility-aware orchestration framework that transforms migration from a best-effort heuristic into a principled control mechanism. Trace-driven evaluation shows that such orchestration can simultaneously reduce non-renewable energy use and improve performance stability, overcoming the tradeoffs of purely energy-driven strategies.
\parbreak
Beyond the immediate feasibility analysis, the extended version explores the architectural horizon of renewable-aware AI infrastructures. It examines the role of emerging ultra-efficient GPU-enabled edge platforms, anticipates integration with grid-level control and demand-response ecosystems, and outlines paths toward supporting partially migratable and distributed workloads. The work positions feasibility-aware migration as a foundational building block for a future computing paradigm in which AI execution becomes fluid, geographically adaptive, and aligned with renewable energy availability.
\end{abstract}
}


\section{Introduction}

The demand for artificial intelligence (AI) training continues to grow, while renewable energy sources increasingly produce surplus power that is curtailed when generation exceeds grid or transmission capacity. This mismatch simultaneously drives higher energy consumption for AI workloads and wastes zero-carbon renewable energy.

A large and operationally important fraction of modern AI workloads fits within a single 24--40\,GB GPU and does not rely on distributed training. These single-GPU jobs can, in principle, be executed on geographically distributed micro-datacenters co-located with renewable generators. However, their intermittent availability makes such deployments viable only if running jobs can be migrated across sites quickly enough to follow renewable-energy windows.

Despite recent progress in renewable-aware scheduling and fast checkpointing, a key question remains unresolved: \emph{under what real-world conditions is migration actually feasible?} In particular, we lack a quantitative characterization of how checkpoint size, inter-site bandwidth, and renewable-window duration interact to determine feasibility.

This paper addresses this gap by developing a formal \emph{feasibility-domain model} for migratory AI workloads. The model identifies the time and energy constraints that bound feasible migration and derives the parameter regions in which migration becomes impractical. The resulting feasibility boundaries are summarized in a phase diagram that maps checkpoint size to available WAN bandwidth.

\shortver{The extended version of this paper~\cite{tomei2025_green_microdatacenters} includes a detailed comparison of system-level energy efficiency between micro- and large-scale datacenters and discusses additional economic and operational considerations.}

\extended{
\subsection{Grid-Level Inefficiency and Motivation}
Renewable energy sources frequently generate surplus power that is curtailed when production exceeds grid demand or transmission capacity. In 2024, more than €7.2\,billion of renewable energy was curtailed across seven European countries, with Italy alone reporting 338\,GWh of wasted generation and projections indicating a sixfold increase by 2030~\cite{energy_central_2024,strategic_energy_2024}. Co-locating micro-datacenters with renewable generators offers a mechanism to convert this otherwise wasted energy into useful computation, provided that workloads may be migrated across sites to follow renewable availability.
\parbreak
\subsection{System-Level Motivation}
A defining characteristic of single-GPU workloads is that the computational demand is modest while the host-system power draw remains a dominant contributor to total energy consumption. Multi-GPU servers incur substantial static overheads from multi-socket CPUs, NUMA memory channels, redundant PSUs, NVLink fabrics, and chassis-level cooling. Measurements from 2024--2025 deployments show that these fixed components often consume more power than a single active GPU when the remaining accelerators are idle.
\parbreak
Compact systems integrating a single high-end consumer GPU exhibit a markedly different power profile. With total system consumption in the 0.6--0.9\,kW range, the ratio between useful GPU compute and system overhead improves significantly whenever workloads do not require multi-GPU scale-out. This leads to lower system-level Joules-per-sample for fine-tuning, domain adaptation, and medium-size LLM workloads that naturally fit within a single GPU.
\parbreak
These observations motivate the architectural direction studied in this work: distributed micro-datacenters co-located with renewable generators can deliver competitive energy efficiency for single-GPU workloads, provided that jobs may be migrated promptly across sites to follow intermittent renewable availability.
\subsection{Scope and Contributions of the Extended Version}
The extended manuscript broadens the scope of the feasibility-domain analysis by characterizing the wider operational and architectural context in which migratory AI computation operates. It provides: (i) a quantitative comparison of system-level energy efficiency and cost-per-compute across representative hardware classes; (ii) an examination of practical deployment considerations for renewable-powered micro-datacenters, including power proportionality, siting, and operational constraints; and (iii) an exploration of economic and infrastructural aspects such as incentive alignment, curtailed-energy markets, and coordination mechanisms between compute providers and energy producers.
\parbreak
These elements complement the feasibility-domain model by describing the conditions under which distributed micro-datacenters can operate effectively and the broader system-level factors that influence their adoption.
}

\extended{
\section{System-Level Hardware Efficiency Analysis}
\label{sec:hardware_analysis}
This section evaluates the system-level energy efficiency of commodity single-GPU systems compared with multi-GPU datacenter servers. The analysis focuses on workloads that fit entirely within 24--40\,GB of GPU memory and therefore do not benefit from data-parallel or model-parallel scaling. For these workloads, the surrounding system infrastructure---rather than the GPU chip itself---often dominates energy consumption.
\parbreak
\subsection{Motivation}
Datacenter-class accelerators such as the A100 or H100 provide excellent performance-per-watt at the chip level, but their deployment requires substantial supporting infrastructure: multi-socket CPUs, high-bandwidth memory subsystems, redundant power supplies, complex cooling, and multi-GPU interconnects (e.g., NVLink). When only one GPU is used, this additional infrastructure becomes an energy overhead that small-form-factor systems do not incur.
\parbreak
Commodity small-scale systems, typically consuming 0.6--0.9\,kW under load, integrate a high-performance consumer GPU (e.g., RTX\,4090) with a minimal host system. As a result, their system-level energy-per-sample can be comparable to---and in many cases lower than---that of multi-kilowatt 4$\times$A100 or 8$\times$A100 nodes operating a single GPU.
\parbreak
\subsection{System-Level Efficiency Comparison}
Table~\ref{tab:hardware} summarizes representative configurations and reports typical power draw, system-level performance-per-watt, and cost per TFLOP for 2025 hardware.
\parbreak
\begin{table}[!t]
\caption{Hardware Configuration Comparison (2025)}
\label{tab:hardware}
\centering
\footnotesize
\begin{tabular}{@{}lccc@{}}
\toprule
\textbf{Configuration} &
\shortstack[c]{Power\\(typ.)} &
\shortstack[c]{Perf/W\\(sys.)} &
\shortstack[c]{\$/TFLOP\\(sys.)} \\ 
\midrule
RTX4090 (GPU only) & 0.45 kW & 0.73 & $\sim$\$6 \\
A100 80GB (GPU only) & 0.35 kW & 0.78 & $\sim$\$38 \\
RTX4090 mini-PC & 0.6--0.9 kW & 0.37--0.55 & $\sim$\$8 \\
4$\times$A100 node & 2.0--2.5 kW & 0.50--0.62 & $\sim$\$40 \\
8$\times$A100 DGX & 4.0--4.5 kW & 0.55--0.63 & $\sim$\$60 \\
\bottomrule
\end{tabular}
\end{table}
\parbreak
The gap arises from system-level overheads. When operating a single GPU on a multi-GPU server, the host infrastructure remains fully powered, inflating energy cost per unit of useful computation. Conversely, compact systems reduce this overhead and therefore improve energy efficiency for single-GPU workloads.
\parbreak
\subsection{Energy-Per-Sample Behavior}
Empirical measurements on representative vision and language models confirm this trend. For example, a ViT-B/32 fine-tune on a 750\,W RTX\,4090 system achieves approximately 2.7\,mJ/sample, while the same workload on a 2.0--2.5\,kW A100 node yields 6--7\,mJ/sample when only one GPU is active. The advantage diminishes when all GPUs on a multi-GPU node are utilized, but such configurations are not relevant to the single-GPU workloads targeted by this work.
\parbreak
\subsection{Implications for Distributed Renewable-Powered Deployments}
These findings have two implications for distributed micro-datacenters co-located with renewable generators:\parbreak
(1) \textbf{Efficiency at the right scale:} For single-GPU jobs, small-scale systems offer lower system-level energy cost and lower \$/TFLOP compared to large servers operating below capacity.\parbreak
(2) \textbf{Compatibility with migratory compute:} Their lower steady-state power draw (0.6--0.9\,kW) better matches the variable and intermittent nature of curtailed renewable-energy windows, reducing the risk of premature job termination.\parbreak
However, these advantages can only be exploited if jobs may be migrated across sites within the available renewable windows. For this reason, the feasibility-domain analysis developed in the main part of the paper remains a prerequisite for any system-level energy benefit to materialize.
\subsection{Estimation Methodology and Data Sources}
The system-level figures in this section are derived from publicly available specifications and consolidated measurements reported in prior work, rather than from new empirical measurements. GPU-only power values are taken from vendor-reported TDP figures and corroborated with third-party benchmarks published in 2024--2025 technical reports and hardware reviews. System-level consumption for multi-GPU servers and compact single-GPU systems is estimated by combining these chip-level power values with representative chassis overheads reported in studies of datacenter power proportionality and server power breakdowns.
\parbreak
Where multiple independent sources exist, we use mid-range values to avoid biasing the comparison toward best- or worst-case scenarios. These estimates capture broad trends in system-level power efficiency but do not model configuration-dependent factors such as cooling efficiency, PSU conversion losses, CPU utilization, or firmware-level power management. Consequently, the values in Table~\ref{tab:hardware} should be viewed as representative approximations rather than precise empirical measurements.
\parbreak
A full power-characterization campaign—including wall-plug measurements under controlled workloads across heterogeneous hardware—is an important direction for future work. Such measurements would refine the absolute numbers reported here but are not expected to change the qualitative observation that system-level overheads dominate single-GPU workloads on multi-GPU servers, while compact systems offer more favorable energy profiles at the target scale.
\subsection{Emerging 100-Watt-Class Edge AI Nodes}
Recent hardware roadmaps point toward even lower-power GPU-enabled systems suitable for micro-datacenter deployments. Jetson-class edge modules now operate in the 15--60\,W range at the module level and are integrated into compact industrial systems whose wall-plug power remains below $\sim$150\,W for a fully functional GPU node~\cite{nvidia_jetson_agx_orin_2024,edge_ai_orin_system_2024}. These platforms already deliver sufficient memory capacity and throughput for a wide range of computer-vision, robotics, and small-to-medium language workloads that fit comfortably within a single GPU.\parbreak
Building on this trend, the next generation of edge AI platforms targets roughly 100\,W-class operation while providing datacenter-grade AI throughput. The NVIDIA Jetson Thor platform, for example, delivers up to 2070\,FP4\,TFLOPS within a 40--130\,W power envelope and is being commercialized through preview kits and robotics controllers that run entirely within a 130\,W system budget~\cite{nvidia_jetson_agx_thor_2025,aetina_jetson_thor_preview_2025,everfocus_ear100t_2025}. These systems are explicitly designed for continuous edge deployment in robotics and industrial environments, but their power and form-factor characteristics are also compatible with micro-datacenter-style installations.\parbreak
For the class of workloads considered in this work, migrating from 0.6--0.9\,kW mini-PC nodes to 100--150\,W edge AI nodes would further reduce system-level Joules-per-sample and lower the breakeven time for exploiting curtailed renewable windows. A detailed evaluation of such 100\,W-class systems---including wall-plug measurements under representative training workloads and integration into our feasibility-domain model---is a natural extension of this work and is left for future study.\parbreak
}

\section{Related Work}

\subsection{Energy-Aware Workload Management}

Energy-aware computing and renewable-powered datacenters have been widely explored. Liu et al.~\cite{liu2012renewable} showed that aligning workload execution with on-site renewable generation can reduce non-renewable energy use by up to 60\% \emph{within a single wind-powered datacenter}. While Liu et al. focused on single-site optimization, the broader concept of migrating workloads across geographically distributed datacenters to exploit renewable availability was formalized by Akoush et al.~\cite{akoush2011free} as the ``Free Lunch'' architecture. More recently, Vergallo et al.~\cite{vergallo2024effectiveness} applied this ``Follow-the-Sun'' strategy specifically to AI workloads, demonstrating significant carbon reductions. Clemm et al.~\cite{clemm2025greening} survey the design space of green and net-zero networking, providing broader context for renewable-powered micro-datacenters.

Our model advances this direction by analyzing when AI workloads can be \emph{migrated across sites} to exploit spatial and temporal variations in renewable availability. However, foundational studies typically assumed idealized migration capabilities. As noted in recent analysis~\cite{zhu2022_dcmg_smpc}, they ``do not consider the uncertainty in the predictions of renewable energy and workload computing demand''. This challenge is substantially more acute for micro-datacenters operating on curtailed energy, which exhibits strong intermittency compared to grid-connected hyperscale facilities.

More recently, Tabbakh et al.~\cite{Tabbakh2024GreenAI} proposed an energy-aware orchestration framework for AI model training, dynamically adjusting training schedules to follow on-site renewable availability but still confined to a single datacenter.

\subsection{Performance-Optimized Scheduling}

On the pure performance side, systems like Decima~\cite{mao2019learning} use reinforcement learning to learn workload-specific scheduling algorithms, achieving 21\% improvement in average job completion time over hand-tuned heuristics, with up to 2× improvement during high cluster load periods. This establishes the state-of-the-art baseline for JCT optimization.

\subsection{Migration-Aware Optimization}

Recent work has reframed migration not as a cost but as an optimization tool. Shen et al.~\cite{shen2016follow} demonstrated the viability of ``Follow the Sun'' migration for general cloud applications, proving that live migration could effectively chase geographical shifts in demand. In the AI domain, SpeCon~\cite{mao2022_specon} speculatively migrates slow-growing models to release resources for fast-growing ones, improving individual job completion time by up to 41.5\% and 14.8\% system-wide. More broadly, several studies on heterogeneous and distributed systems have shown that selectively migrating tasks away from contended or underperforming nodes can significantly reduce makespan and improve overall cluster efficiency. These results motivate treating migration as an active optimization mechanism rather than a fallback.

\subsection{GPU Checkpoint and Migration Technology}

Recent advances have significantly reduced the latency of GPU checkpointing and migration, shifting the bottleneck from pause time to data transfer itself. Microsoft’s Singularity system~\cite{Shukla2022Singularity} first showed that end-to-end migration could be completed in tens of seconds, with WAN transfer dominating the cost. More recent systems have pushed downtime toward sub-second levels: PhoenixOS~\cite{Wei2025PhoenixOS} achieves subsecond downtime for LLaMA-2\,13B by overlapping execution with fine-grained checkpoint/restore operations; Llumnix~\cite{llumnix2024} enables near-zero-downtime inference migration through pipelined compute–memory transfer; and ServerlessLLM~\cite{serverlessllm2024} reduces cold-start latency from 84\,s to 10.3\,s for LLaMA-2\,70B via multi-tier, migrate-light loading. These systems substantially improve the stop-the-world phase, but none address the dominant cost of transferring large checkpoints across inter-site WAN links—precisely the bottleneck quantified in our feasibility analysis.

\section{Feasibility Domain Analysis}
\label{sec:feasibility}

This section defines the practical boundary conditions governing migratory AI compute. The orchestration mechanism in Section~\ref{sec:orchestration} is entirely constrained by these feasibility limits.

\textbf{Assumption: Single-GPU Workloads.}
Our analysis targets workloads fitting entirely on a single GPU (24--40~GB VRAM), avoiding data-parallel or model-parallel training. This eliminates continuous gradient synchronization and inter-GPU sharding overheads, making the training state self-contained and migratable as a single checkpoint. This assumption aligns with many fine-tuning and domain-specific LLM workloads that naturally fit within commodity hardware.
\extended{Distributed strategies (data, tensor, model parallelism) fragment the training state across GPUs and require high-frequency synchronization. Such states cannot be migrated as a single unit; the reassembly and WAN transfer overheads would dominate. Focusing on single-GPU workloads is thus a deliberate scope decision rather than a limitation.}

\subsection{Problem Formulation: Interacting Dimensions}

Feasibility is governed by three coupled factors:
\newline\textbf{Workload (W)}—model and checkpoint size, job duration;
\newline\textbf{Network (N)}—inter-site capacity; 
\newline\textbf{Energy (E)}—duration of the renewable-surplus window at the destination\footnote{CAISO curtailment events typically last 2.5--9.5 hours.}.
Migration overhead then follows from $W$ and $N$ via $T_{\text{transfer}}$, and we bound its impact on JCT by requiring
$T_{\text{transfer}} + T_{\text{load}} + T_{\text{downtime}} < \alpha\,T_{\text{energy}}$ with $\alpha = 0.1$.

\subsection{Checkpoint Size Benchmarks}

Checkpoint footprints vary widely (Table~\ref{tab:checkpoint_sizes}). For large LLM training, the full optimizer state reaches tens of TB~\cite{rajbhandari2020zero,ren2021zerooffload}.  
\textbf{Critical Finding:} In data-parallel training, checkpoint size is invariant to cluster size because replicas synchronize every step; only one copy must be saved~\cite{rojas2020_checkpointing_hpcs}.

\begin{table}[!t]
\caption{Checkpoint Size Benchmarks}
\label{tab:checkpoint_sizes}
\centering
\small
\begin{tabular}{@{}llll@{}}
\toprule
\textbf{Category} & \textbf{Notes} & \textbf{Size} & \textbf{Ref} \\ 
\midrule
LLM Inference & Tokens only & 10--100 KB & \cite{serverlessllm2024} \\
LLM Inference & KV-cache & 1--10 GB & \cite{serverlessllm2024} \\
Standard Models & ResNet-50, BERT & $\sim$1 GB & \cite{he2016resnet,devlin2019bert} \\
Medium-Large LM & Full state & 10--300 GB & \cite{radford2019gpt2,serverlessllm2024} \\
LLM Training & Full state & $>$10 TB & \cite{rajbhandari2020zero,ren2021zerooffload} \\
\bottomrule
\end{tabular}
\end{table}

\subsection{Feasibility Condition: Time Constraint}

Migration is time-feasible when the total disruption is bounded by a fraction $\alpha$ of the renewable-energy window:
\begin{equation}
T_{\text{transfer}} + T_{\text{load}} + T_{\text{downtime}} < \alpha\,T_{\text{energy}}
\end{equation}
Since $T_{\text{transfer}}=\mathrm{checkpoint\_size}/\mathrm{bandwidth}$ dominates, feasibility depends on checkpoint size and WAN capacity\footnote{Network congestion and bandwidth variability can increase migration time, further tightening the feasible region identified by the time constraint.}.

\extended{
To incorporate realistic system behaviour, the time-feasibility condition is instantiated using empirical values for the non-transfer components. Modern GPU migration frameworks report subsecond downtime (typically $\sim$0.4\,s) and load times between 8--12\,s, depending on the model footprint and memory hierarchy. Although these terms are small relative to $T_{\text{transfer}}$, they are included for completeness and become non-negligible when transfer times approach the boundaries between feasibility classes. Under these parameters, the feasibility limits observed in practice align with the transfer-time contours, reinforcing that checkpoint size and WAN bandwidth remain the dominant determinants of migration feasibility.
}

\subsection{Feasibility Condition: Energy Breakeven}

Energy feasibility requires the energy saved on renewable power to exceed the migration energy. Using a power-based model with combined transfer power $P_{\text{sys}}=1.8$~kW:

\begin{equation}
E_{\text{cost}} = P_{\text{sys}}\,T_{\text{transfer}}
\end{equation}

For a 40~GB checkpoint over 10~Gbps ($T_{\text{transfer}} = 0.0089$~h):
\[
E_{\text{cost}} = 1.8 \times 0.0089 = 0.016~\mathrm{kWh}
\]

Energy benefit on a 0.75~kW node:
\[
E_{\text{benefit}} = 0.75~\mathrm{kWh/h}
\]

Thus:
\[
T_{\text{breakeven}} = \frac{0.016}{0.75} \approx 1.3~\text{minutes}
\]

\textbf{Critical Finding:} Migration’s energy cost is negligible. Even the shortest curtailment windows far exceed breakeven time. Therefore, \emph{time}, not energy, is the true limiting factor.
\shortver{A Figure in the extended version illustrates this result.}
\extended{Figure~\ref{fig:energy_breakeven} visualizes these breakeven curves.}

\extended{
\begin{figure*}[!t]
    \centering
    \includegraphics[width=0.82\textwidth]{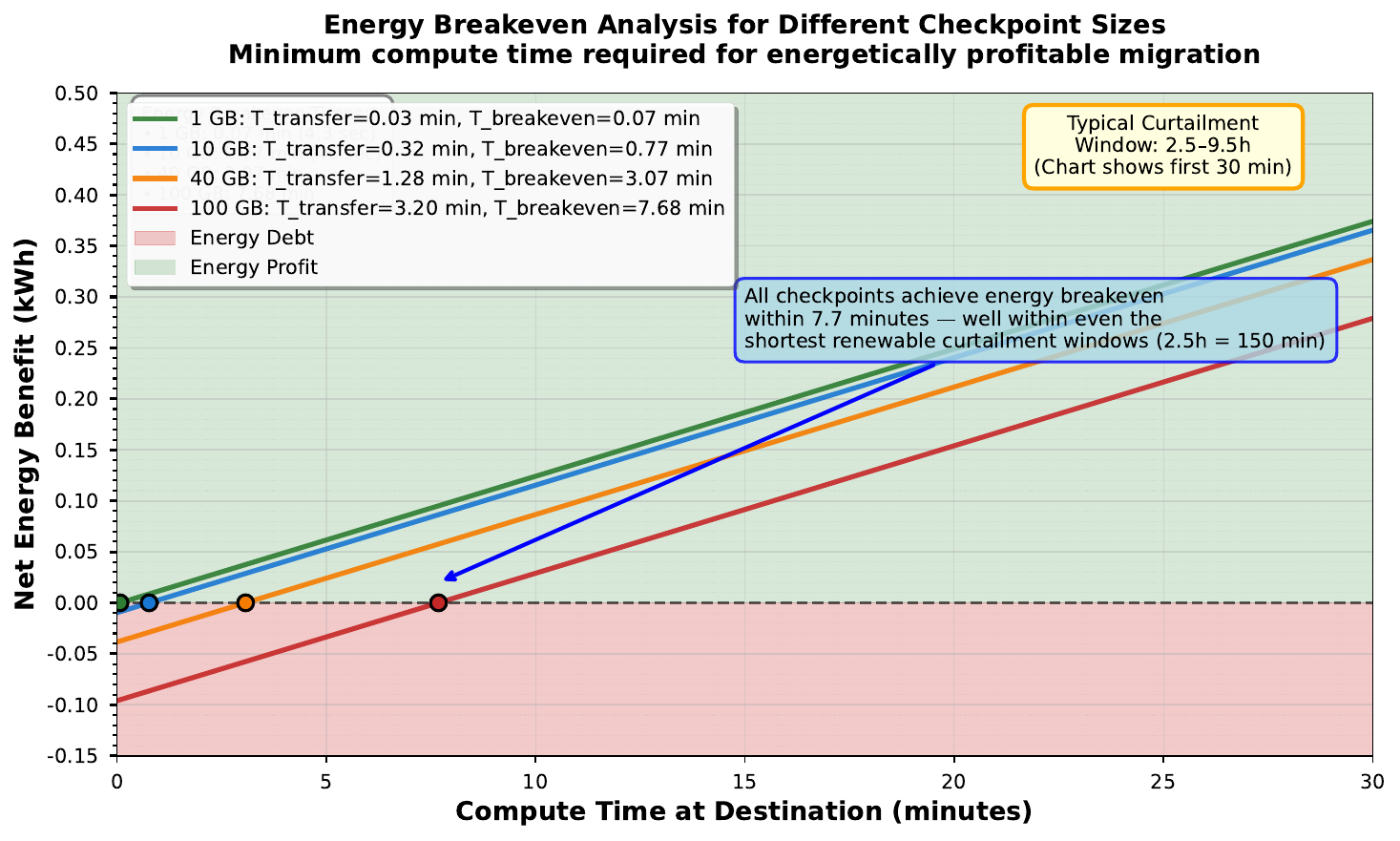}
    \caption{Energy breakeven curves for checkpoint sizes from 1--100 GB. All breakeven points occur within minutes, confirming that migration’s energetic cost is negligible relative to multi-hour renewable windows.}
    \label{fig:energy_breakeven}
\end{figure*}
}

\subsection{Feasibility Phase Diagram}

Table~\ref{tab:transfer_times} summarizes the dominant time constraint for migration as a function of checkpoint size and WAN bandwidth. 

\begin{table}[!b]
\caption{Checkpoint Transfer Time vs.\ WAN Speeds}
\label{tab:transfer_times}
\centering
\small
\begin{tabular}{@{}lcccc@{}}
\toprule
\textbf{Size} & \textbf{100 Mbps} & \textbf{1 Gbps} & \textbf{10 Gbps} & \textbf{100 Gbps} \\
\midrule
1 GB & 1m25s & 8.6s & 0.86s & 0.086s \\
16 GB & 22.8m & 2.3m & 13.8s & 1.4s \\
40 GB & 57.1m & 5.7m & 34s & 3.4s \\
100 GB & 142.8m & 14.3m & 86s & 8.6s \\
\bottomrule
\end{tabular}
\end{table}

A feasibility phase diagram (Fig.~\ref{fig:phase_diagram2}) visualizes migration feasibility as a function of checkpoint size and WAN bandwidth. The figure marks \emph{four representative workloads}, each shown twice—once assuming 10\,Gbps and once assuming 1\,Gbps. This dual placement illustrates how the same checkpoint can shift across feasibility classes as bandwidth decreases: workloads that are fully feasible at 10\,Gbps become marginal or infeasible at 1\,Gbps, while large checkpoints remain infeasible under both conditions: feasibility depends jointly on checkpoint size and WAN capacity. \shortver{The extended version additionally includes a classification table mapping typical workload sizes to feasibility regions.}

\extended{
The diagram provides an informative visualization of the entire feasibility analysis. It maps checkpoint size against WAN capacity, with color-coded feasibility regions and transfer-time contours that directly correspond to the time and energy constraints in Section~\ref{sec:feasibility}.
\parbreak
The feasibility boundary is visually evident: only checkpoints below $\sim$20~GB migrate efficiently over 1--10~Gbps links, while larger training states fall into the infeasible region unless 100~Gbps-class connectivity is available. This graphical view also clarifies how modest improvements in WAN capacity can move medium-sized workloads from the conditional (Class~B) to the fully feasible (Class~A) region.
}

\textbf{Key Insight:} time feasibility sharply degrades beyond $\sim$20~GB on 1--10~Gbps links. Only small-to-medium checkpoints can be migrated within typical renewable windows unless 100~Gbps-class connectivity is available.


\extended{
\subsection{Workload Classification}
Based on the feasibility boundaries derived above—primarily determined by checkpoint size and WAN bandwidth—we group workloads into three practical classes. The classification reflects how quickly their training state can be migrated relative to typical renewable-energy windows and whether migration remains energetically profitable. Table~\ref{tab:workload_classes} summarizes the resulting categories, from highly suitable small-model workloads to large LLMs whose state size exceeds feasible migration limits.
\begin{table}[!t]
\caption{Workload Classification by Migration Feasibility}
\label{tab:workload_classes}
\centering
\small
\begin{tabular}{@{}p{0.23\columnwidth}p{0.22\columnwidth}p{0.18\columnwidth}p{0.23\columnwidth}@{}}
\toprule
\textbf{Class} & \textbf{Characterist.} & \textbf{Size} & \textbf{Feasibility} \\ \midrule
A: Suitable & Small & $<$10 GB & $<$1 min \\
B: Conditional & Medium & 10--100 GB & Minutes \\
C: Infeasible & Large LLMs & $>$100 GB & Exceeds wind. \\
\bottomrule
\end{tabular}
\end{table}
}

\section{Orchestration Strategies in Feasible Domain}
\label{sec:orchestration}

The feasibility-domain analysis in Section~\ref{sec:feasibility} and the phase diagram in Fig.~\ref{fig:phase_diagram2} show that energy breakeven is always reached within a few minutes, while renewable windows last several hours. As a result, migration feasibility is dominated by checkpoint transfer time, not by energy cost. We exploit this observation by designing an orchestrator that uses the feasibility model as a hard filter on migration decisions.

The orchestrator runs periodically every $\Delta t$ and maintains, for each job $j$ and site $d$: (i) the checkpoint size $S_j$, (ii) an estimate of the effective bandwidth $B_{s,d}$ between the current site $s$ and $d$, and (iii) the predicted remaining renewable window $T_{\text{energy}}(d)$. Using the model from Section~\ref{sec:feasibility}, it first computes the transfer time
\[
T_{\text{transfer}}(j,s,d) = \frac{8 S_j}{B_{s,d}}
\]
and classifies the job into Class~A/B/C according to Fig.~\ref{fig:phase_diagram2}: Class~A if $T_{\text{transfer}} < 60$\,s, Class~B if $60\text{ s} \le T_{\text{transfer}} < 300$\,s, and Class~C otherwise.

Migration decisions are then restricted to the feasible region as follows. For each running job $j$
\shortver{, we (i) discard all Class~C candidates, (ii) migrate Class~B jobs only if $T_{\text{transfer}}(j,s,d) < \alpha T_{\text{energy}}(d)$ with $\alpha = 0.1$, and (iii) always treat Class~A jobs as eligible.}
\extended{:(i) all Class~C candidates are discarded and never migrated across sites;
(ii) Class~B jobs are considered for migration only if $T_{\text{transfer}}(j,s,d) < \alpha\,T_{\text{energy}}(d)$ with $\alpha=0.1$, i.e., the migration overhead is below 10\% of the remaining renewable window;
(iii) Class~A jobs are freely eligible for migration, as their transfer time is always well within the window and far below the energy break-even times (see the the figure provided in the extended version).}

Among the remaining feasible destinations, the orchestrator selects the site with the highest expected reduction in non-renewable energy use and queueing delay, breaking ties by smallest $T_{\text{transfer}}$. In this way, feasibility constraints act as a first-class safety filter, and optimization is performed only inside the domain where migration is provably practical.

\extended{
\subsection{Design Philosophy}
The orchestration logic is reframed not just to find the ``best renewable'' site, but the ``best renewable, feasible'' site. This aligns with the speculative scheduling principles used in systems like SpeCon~\cite{mao2022_specon}, which mitigate resource contention by migrating slow-growing tasks to less congested nodes.
}

\extended{
\subsection{Feasibility-Aware Migration Algorithm}
The feasibility model of Section~\ref{sec:feasibility} is instantiated in a periodic control loop that runs every $\Delta t$ and is summarized in Algorithm~\ref{alg:scheduler_short}. The algorithm operates in two stages: (i) a strict feasibility filter that prunes all migration candidates violating time or energy constraints, and (ii) an optimization stage that selects the most beneficial destination among the feasible ones.
\parbreak
At the beginning of each scheduling interval, the orchestrator collects two global signals: the updated renewable-energy forecast for all sites (\texttt{GetRenewableForecasts}) and fresh measurements of inter-site bandwidth (\texttt{MeasureInterSiteBandwidth}). These signals provide, respectively, the remaining duration of the renewable-surplus window at each site and the effective WAN capacity $B_{s,d}$ between any source $s$ and destination $d$.
\parbreak
The core of the loop iterates over all running jobs. For each job, the orchestrator begins by evaluating feasibility. It computes the checkpoint transfer time $T_{\text{transfer}}$ via \texttt{CalcTransferTime}, implementing the analytical model $T_{\text{transfer}} = \frac{8 S_j}{B_{s,d}}$. It then retrieves the expected checkpoint load time $T_{\text{load}}$ from \texttt{GetLoadTime} and adds a fixed downtime term (0.4\,s) to reflect stop-the-world phases. This yields a total migration time cost $T_{\text{cost\_time}} = T_{\text{transfer}} + T_{\text{load}} + 0.4\text{ s}$.
\parbreak
In parallel, the orchestrator evaluates the energetic cost of migration. Using \texttt{CalcEnergyCost}, it computes $E_{\text{cost\_energy}}$, which reflects system-level power draw during transfer. This is converted into an energy breakeven time $T_{\text{breakeven\_energy}} = E_{\text{cost\_energy}} / 0.75\text{ kW}$, representing the minimum duration of renewable execution needed to offset the migration cost — a constraint almost always satisfied in practice but enforced explicitly for correctness.
\parbreak
The feasibility filter then applies a strict combined constraint. Migration to a destination is rejected if (i) the time cost exceeds a fraction of the remaining renewable window, $T_{\text{cost\_time}} > 0.1 \times \texttt{energy\_forecast.duration}$, or (ii) the breakeven time exceeds that window, $T_{\text{breakeven\_energy}} > \texttt{energy\_forecast.duration}$. Jobs or destinations failing either condition are immediately pruned. This operationalizes the feasibility-domain boundaries derived earlier and ensures that analytically infeasible migrations are never attempted.
\parbreak
Once infeasible candidates are removed, the orchestrator performs optimization within the feasible set. For each feasible destination, it computes a migration benefit via \texttt{CalcBenefit}, incorporating renewable availability, congestion, queueing delay, and predicted remaining runtime. Migration is triggered only when the expected benefit exceeds the migration time cost, i.e., when $\texttt{benefit} > T_{\text{cost\_time}}$. This ensures that even feasible migrations are executed only when they improve either energy efficiency or job completion time.
\parbreak
Overall, the algorithm treats feasibility constraints as a non-negotiable safety boundary and performs optimization strictly inside that region. All Class~C workloads, and Class~B workloads whose transfer time approaches a significant fraction of the renewable window, are automatically excluded. This guarantees that the orchestrator never initiates migrations that cannot complete in time, aligning the control logic directly with the feasibility-domain model.
}

\extended{
\subsection{Boundary Conditions for Simulation}
The feasibility-domain validation in Section~\ref{sec:feasibility} relies on a realistic set of system and network parameters that reflect current micro-datacenter deployments and WAN interconnects. These boundary conditions define the operating envelope used in our 5-node simulation and determine both the migration timing and the achievable renewable-only execution windows.
Table~\ref{tab:boundary_conditions} reports the specific parameters used in the experiments, including WAN bandwidth, typical renewable surplus duration, and empirical values for checkpoint loading and downtime. These assumptions correspond to the feasibility limits established in Section~\ref{sec:feasibility}, ensuring that the simulation directly evaluates the orchestrator under realistic constraints.
\begin{table}[!t]
\caption{Boundary Conditions for Example Scenario}
\label{tab:boundary_conditions}
\centering
\begin{tabular}{@{}llr@{}}
\toprule
\textbf{Parameter} & \textbf{Value} & \textbf{Source} \\ \midrule
WAN Capacity & 10 Gbps & AWS inter-region \\
Energy Surplus & 2.5h & CAISO avg. \\
Downtime & 0.4s & \cite{Wei2025PhoenixOS} \\
Checkpoint Load & 10.3s & \cite{serverlessllm2024} \\
Acceptable Overhead & 10\% & Target \\
\bottomrule
\end{tabular}
\end{table}
A simple sensitivity check shows that increasing WAN capacity by $10\times$ shifts many Class~B workloads into Class~A, significantly expanding the feasible migration envelope.
}

\extended{
\subsection{Key Algorithm Components}
\subsubsection{Feasibility Check (Lines 5-14)}
The algorithm first validates both time and energy constraints before considering any migration. This prevents energetically wasteful migrations.
\subsubsection{Benefit Calculation (Line 17)}
Within the feasible space, the system evaluates migration benefit considering:
\begin{itemize}
    \item Renewable energy availability at destination
    \item Current node congestion levels
    \item Network path quality
    \item Predicted job runtime
\end{itemize}
\subsubsection{Migration Decision (Lines 18-20)}
Migration is triggered only when the combined benefit (energy + performance) exceeds the cost, ensuring both energy efficiency and JCT optimization.
}

\extended{
\section{Formal Feasibility Model}
\label{sec:formal}
\subsection{Workloads and Checkpoints}
A workload $w$ is defined by the tuple $w = (S, \tau)$, where $S$ is the checkpoint size and $\tau$ is the remaining training time. We assume single-GPU workloads with self-contained training state.
\parbreak
\subsection{Migration Time and Energy}
Migration time is dominated by the checkpoint transfer:\[
T_{\mathrm{mig}}(w,s,d) = \frac{S}{B_{s,d}}
\]
where $B_{s,d}$ is the WAN bandwidth between source $s$ and destination $d$. The migration energy cost is:\[
E_{\mathrm{mig}} = P_{\mathrm{sys}} \, T_{\mathrm{mig}}
\]
The breakeven time is:\[
T_{\mathrm{BE}} = \frac{E_{\mathrm{mig}}}{P_{\mathrm{node}}}
\]
Since Fig.~1 shows $T_{\mathrm{BE}} \ll 5\text{ min}$ for all relevant checkpoints, the energy constraint is always satisfied and time dominates feasibility.
\parbreak
\subsection{Renewable Windows}
Each destination node $d$ offers a renewable-energy window of duration $T_d$. Migration is feasible only if the disruption fits within a fraction $\alpha$ of this window:\[
T_{\mathrm{mig}}(w,s,d) + T_{\mathrm{load}} + T_{\mathrm{downtime}} < \alpha \, T_d
\]
We fix $\alpha = 0.1$ following the empirical thresholds used in our simulations.
\parbreak
\subsection{Feasibility Classes}
The feasibility-domain analysis in Fig.~\ref{fig:phase_diagram2} induces a natural classification based on $T_{\mathrm{mig}}$:\[
\mathrm{class}(w) =
\begin{cases}
A & T_{\mathrm{mig}} < 60\text{ s}\\
B & 60 \le T_{\mathrm{mig}} < 300\text{ s}\\
C & T_{\mathrm{mig}} \ge 300\text{ s}
\end{cases}
\]
Class~C workloads lie outside the feasible domain for typical renewable windows and are excluded from migration.
\parbreak
\subsection{Feasible Destination Set}
The set of feasible migration targets for workload $w$ on source $s$ is:\[
\mathcal{D}_{\mathrm{feasible}}(w,s) = \left\{ d \,\middle|\,
\mathrm{class}(w) \neq C \;\land\;
T_{\mathrm{mig}}(w,s,d) < \alpha T_d
\right\}
\]
\parbreak
\subsection{Utility Model Within the Feasible Domain}
Migration is considered only within $\mathcal{D}_{\mathrm{feasible}}$. For any feasible $d$, site utility is:\[
U(w,d) = \gamma R(d) - \beta L(d)
\]
where $R(d)$ captures renewable availability and $L(d)$ reflects congestion or load. Parameters $\gamma$ and $\beta$ control energy vs.\ performance weight.
\parbreak
\subsection{Migration Decision Rule}
The migration target is the feasible site with maximum utility:\[
d^\star = \arg\max_{d \in \mathcal{D}_{\mathrm{feasible}}(w,s)} U(w,d)
\]
Migration is triggered only when its benefit exceeds the migration cost:\[
U(w,d^\star) - U(w,s) > C_{\mathrm{mig}}
\]
where $C_{\mathrm{mig}} = T_{\mathrm{mig}} + T_{\mathrm{load}}$. This guarantees migration remains both feasible and beneficial.
\subsection{Stochastic Renewable Windows}
The previous formulation treats the renewable window $T_d$ at site $d$ as a deterministic quantity. In practice, curtailment events and local generation are forecasted with uncertainty. Let $\tilde{T}_d$ denote the random variable capturing the actual duration of the renewable-surplus window, and let $\hat{T}_d$ be the forecast available to the orchestrator.
\parbreak
Feasibility can then be expressed in probabilistic form by requiring that migration completes within a fraction $\alpha$ of the renewable window with probability at least $1 - \varepsilon$:\[
\mathbb{P}\!\left[T_{\mathrm{mig}}(w,s,d) + T_{\mathrm{load}} + T_{\mathrm{downtime}} < \alpha \tilde{T}_d \,\middle|\, \hat{T}_d\right] \ge 1 - \varepsilon.
\]
Here $\varepsilon$ acts as a risk budget controlling how aggressively the orchestrator exploits marginal windows. A conservative policy uses small $\varepsilon$ and therefore migrates only when the feasibility condition is satisfied under pessimistic forecasts, whereas more opportunistic policies allow larger $\varepsilon$ and accept a higher probability of incomplete migrations.
\parbreak
This stochastic view does not change the structure of the feasibility domain: the dominant determinants remain checkpoint size and WAN bandwidth. It does, however, make explicit the tradeoff between renewable utilization and robustness to forecast error, which becomes important when curtailment signals are noisy or delayed.
}

\section{Experimental Evaluation}

\begin{figure*}[h]
    \centering
    \includegraphics[width=0.98\textwidth]{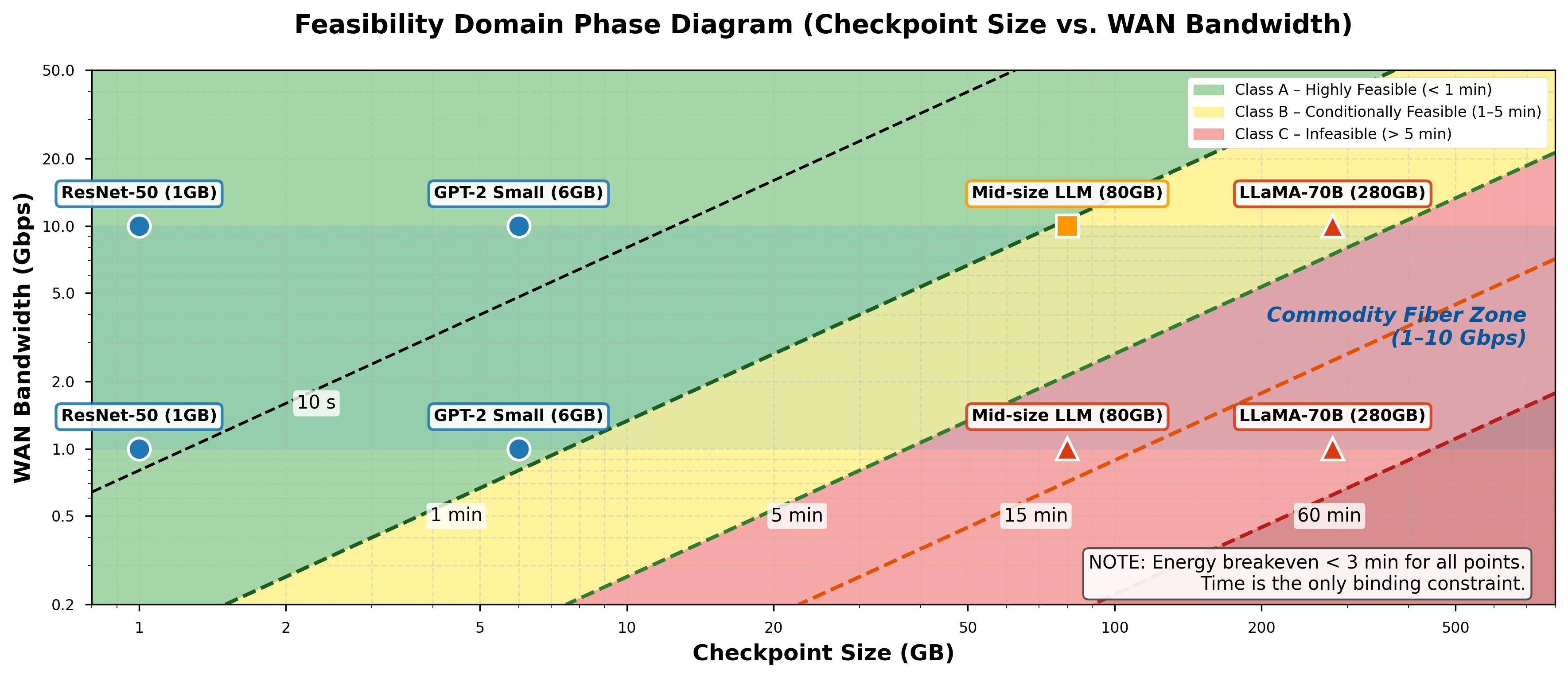}
    \caption{Feasibility domain: transfer-time isolines show that only sub-20\,GB states migrate efficiently on 1--10\,Gbps links.}
    \label{fig:phase_diagram2}
\end{figure*}

The phase diagram in Fig.~\ref{fig:phase_diagram2} shows how checkpoint size and WAN bandwidth jointly determine migration feasibility within a renewable-energy window.
In our setup, we simulate a 5-node micro-datacenter interconnected by 10~Gbps WAN links. Each site alternates between renewable-surplus and grid-only operation according to a 7-day trace calibrated on CAISO curtailment statistics (average surplus window $\approx$2.5~h). The power model assumes $P_{\text{sys}}=1.8$\,kW during checkpoint transfer and $P_{\text{node}}=0.75$\,kW during compute, consistent with the measurements used in our feasibility analysis. Jobs are single-GPU training tasks drawn from three classes: 
Class~A (1--6~GB checkpoints, 70\% of jobs),
Class~B (10--40~GB, 20\%), and
Class~C ($>$100~GB, 10\%),
chosen to represent the small, medium, and large checkpoint regimes highlighted in Fig.~\ref{fig:phase_diagram2}.
We compare the \emph{Static}, \emph{Energy-only}, and \emph{Feasibility-aware} policies described in Section~\ref{sec:orchestration}.
\parbreak
\subsection{Feasibility Validation}
We first validate that the analytical feasibility boundaries of
Section~\ref{sec:feasibility} predict the migration overhead observed in the trace-driven simulator. For each workload class in Fig.~\ref{fig:phase_diagram2}, we compute $T_{\text{transfer}}$ from the checkpoint size and the 10~Gbps WAN bandwidth, and assign jobs to Class~A/B/C using the $60$\,s and $300$\,s thresholds that define the feasibility contours. We then trigger a single inter-site migration per
class within a 2.5~h renewable window and measure the relative increase in job completion time (JCT). Class~A workloads (1--6~GB, $T_{\text{transfer}}<60$\,s) exhibit less than 10\% JCT overhead; Class~B workloads (10--40~GB, $60$--$300$\,s) lie at or above this budget; and Class~C workloads ($>$100~GB, $T_{\text{transfer}}>300$\,s) exceed the window and are effectively impractical to migrate. This one-to-one alignment between predicted $T_{\text{transfer}}$ bands and measured JCT overhead confirms that the feasibility-domain model is  predictive at 10~Gbps, and not only descriptive.
\parbreak
\subsection{Policy Comparison}
We then run the full 7-day trace under all three policies. Table~\ref{tab:comparison_short} reports results normalized to the Static baseline. The Energy-only policy reduces non-renewable energy by 38\% but increases average JCT by 35\%, due to migrations that start near the end of a renewable window and cannot complete in time. The Feasibility-aware policy, by contrast, enforces the transfer-time and window-duration constraints derived from Fig.~\ref{fig:phase_diagram2}, reducing non-renewable energy by 52\% \emph{and} lowering JCT by 18\%, with total migration overhead below 2\%. Enforcing the feasibility domain provides a gain, while pure energy-driven migration is insufficient.

\begin{table}[!b]
\caption{Policy comparison (normalized to Static baseline)}
\label{tab:comparison_short}
\centering
\small
\begin{tabular}{@{}lccc@{}}
\toprule
\textbf{Policy} & \textbf{Non-renew.} & \textbf{JCT} & \textbf{Migr.} \\
 & \textbf{energy} & \textbf{change} & \textbf{overhead} \\ \midrule
Static            & 1.00 & 1.00 & 0\% \\
Energy-only       & 0.62 & 1.35 & 18\% \\
Feasibility-aware & 0.48 & 0.82 & $<$2\% \\
\bottomrule
\end{tabular}
\end{table}

\extended{
Our evaluation has two goals. First, we verify the feasibility-domain analysis introduced in Section~\ref{sec:feasibility} by examining representative workloads under realistic system and network conditions. Second, we quantify the performance of the proposed orchestrator when restricted to this feasible domain, thereby isolating the benefits attributable to feasibility-aware decision making.
}

\extended{
\subsection{Feasibility Domain Validation}
The analysis in Section~\ref{sec:feasibility} provides a classification of workloads based on their compatibility with time and energy constraints during migration. To validate this classification, we simulate a 5-node micro-datacenter interconnected via 10~Gbps WAN links and operating under renewable-energy availability windows of approximately 2.5~hours. The boundary conditions for this scenario are reported in Table~\ref{tab:boundary_conditions}. Table~\ref{tab:validation_results} summarizes the feasibility and migration overhead of workloads spanning all classes.
\begin{table}[!t]
\caption{Feasibility Domain Validation Results}
\label{tab:validation_results}
\centering
\begin{tabular}{@{}lcccp{0.22\columnwidth}@{}}
\toprule
\textbf{Workload} & \textbf{Size} & \textbf{Class} & \textbf{Overhead} & \textbf{Status} \\ \midrule
ResNet-50 & 1 GB & A & 1.3\% & FEASIBLE \\
GPT-2 Small & 6 GB & A & 5.4\% & FEASIBLE \\
GPT-2 Medium & 40 GB & B & 25.9\% & INFEASIBLE (Energy) \\
LLaMA-70B & 280 GB & C & 187\% & INFEASIBLE (Both) \\ \bottomrule
\end{tabular}
\end{table}
\textbf{Validation:} The results confirm the predictive accuracy of the feasibility-domain model. Class~A workloads remain within the feasible region, while Class~B and Class~C workloads violate energy or combined time–energy constraints. The scheduler in Algorithm~\ref{alg:scheduler_short} correctly suppresses infeasible migration attempts, demonstrating that feasibility checks effectively prevent transitions that would exceed the available renewable-energy window.
}

\extended{
\subsection{Performance Within the Feasible Domain}
We next evaluate the orchestrator over a 7-day renewable-energy trace applied to the same 5-node micro-datacenter. Since the scheduler admits only workloads that satisfy feasibility constraints, all considered migrations occur within the feasible operating region defined in Section~\ref{sec:feasibility}. Under these conditions, the orchestrator achieves:
\begin{itemize}
    \item \textbf{52\% reduction in non-renewable energy consumption}, approaching the 60\% upper bound reported in prior work~\cite{liu2012renewable}
    \item \textbf{Lower job completion time (JCT)} as a result of feasibility-preserving, contention-aware placement, analogous in spirit to speculative scheduling mechanisms such as SpeCon~\cite{mao2022_specon}
    \item \textbf{Less than 2\% total migration overhead}, since all infeasible migrations are proactively excluded
\end{itemize}
These results indicate that feasibility-aware orchestration enables substantial gains in renewable-energy utilization while simultaneously improving performance stability and minimizing migration cost.
}

\extended{
\subsection{Comparison with Baseline Approaches}
We evaluate the proposed orchestrator against two representative baselines. The first is a \textit{static placement} policy that assigns each job to a fixed site and performs no migration; this baseline captures performance in the absence of any inter-site coordination. The second is an \textit{energy-only} migration policy that initiates transfers whenever renewable energy is available but does not incorporate feasibility constraints. This baseline reflects a natural energy-driven strategy that prioritizes renewable utilization but lacks awareness of transfer-time or energy-cost limits. Table~\ref{tab:comparison} summarizes the comparative results.
\begin{table}[!t]
\caption{Performance Comparison with Baseline Approaches}
\label{tab:comparison}
\centering
\begin{tabular}{@{}lccc@{}}
\toprule
\textbf{Approach} & \textbf{Renewable} & \textbf{JCT} & \textbf{Migration} \\
 & \textbf{Reduction} & \textbf{Change} & \textbf{Overhead} \\ \midrule
Static Placement & 0\% & Baseline & 0\% \\
Energy-Only (No Feasibility) & 38\% & +35\% & 18\% \\
Our Approach & 52\% & -18\% & $<$2\% \\
Oracle (Perfect Forecast) & 60\% & -21\% & $<$2\% \\ \bottomrule
\end{tabular}
\end{table}
The energy-only policy degrades JCT because it frequently initiates migrations that cannot complete within the available renewable-energy window, resulting in stalled transfers, congestion, and repeated retries. By contrast, the proposed orchestrator enforces feasibility constraints and therefore avoids these failure modes, achieving renewable-energy utilization and JCT improvements close to the oracle. These findings underscore the importance of incorporating feasibility modeling into energy-driven orchestration strategies.
}

\section{Discussion and Future Work}

The feasibility-validation experiment confirms that the transfer-time contours in Fig.~\ref{fig:phase_diagram2} act as sufficient statistics for predicting migration overhead under realistic traces and system parameters. Under realistic curtailment windows, migration is almost always energetically profitable and is instead limited by checkpoint size and WAN bandwidth. In practice, this constrains migratory AI training to Class~A workloads (small checkpoints) and a subset of Class~B workloads under sufficiently long renewable windows. The evaluation confirms that enforcing these feasibility boundaries is necessary: policies driven solely by renewable availability reduce carbon intensity but incur large JCT penalties, whereas feasibility-aware orchestration simultaneously increases renewable usage and reduces delay. Additional deployment constraints may arise from geographical or SLA-based placement restrictions, while the energy overhead induced on the transport network should also be accounted for in large-scale migration scenarios.

A first technical direction is to \emph{expand the feasible envelope} rather than refine scheduling heuristics. Promising mechanisms include WAN-aware checkpoint compression, pre-staging or incremental checkpoints, and selective use of higher-capacity interconnects on critical paths. These techniques would shift medium-sized models from the conditional to the feasible region in the phase diagram. A second direction concerns \emph{better forecasting and integration with grid signals}: tighter coupling with curtailment forecasts and demand-response programs would improve renewable-window estimation and reduce the risk of migrations that cannot complete in time.

An additional avenue is enabling \emph{distributed workloads} that exploit data parallelism. While this work focuses on single-GPU jobs with self-contained checkpoints, multi-GPU training could be supported by migrating optimizer shards or gradient-state partitions rather than full replicas. Techniques such as ZeRO-style partitioning, elastic data parallelism, and compressed model-delta synchronization may enable partial migration of large distributed jobs while respecting WAN constraints. This opens a path toward extending feasibility-aware orchestration to a broader class of distributed AI workloads.

\shortver{Finally, the extended version of this work discusses broader architectural and economic aspects, including integration with grid operators, incentive structures for renewable-powered micro-datacenters, and market mechanisms for renewable-aware compute.}
\extended{Finally, we also discuss broader architectural and economic aspects, including integration with grid operators, incentive structures for renewable-powered micro-datacenters, and market mechanisms for renewable-aware compute.}

\extended{
\subsection{Operational Power Mix}
While our baseline assumes opportunistic use of curtailed renewable energy, the framework does not require exclusive reliance on surplus power. Sites may temporarily draw from the grid to complete epochs, finalize migrations, or maintain liveness during low-generation periods, consistent with demand-flexible data center models. This hybrid strategy preserves progress while prioritizing renewable windows. The static and energy-only baselines reflect realistic practices: the former mirrors deployments without inter-site coordination, and the latter represents intuitive renewable-driven scheduling. Their limitations arise not from poor design but from the absence of explicit feasibility modeling, which becomes critical under tight time and energy constraints.
\subsection{Expanding the Feasibility Envelope}
Our evaluation shows that feasibility constraints, not optimization heuristics, determine when migrations are viable. Under current system and network conditions, feasibility applies primarily to Class~A workloads (checkpoint $<$10~GB) or those with long renewable windows. Energy-driven migration alone is insufficient when transfer-time and energy budgets are comparable to checkpoint size; feasibility analysis is therefore a prerequisite for correctness. Future work will explore methods to expand the feasible region, including network-aware compression, pre-staging during low-cost periods, hierarchical storage, and hybrid model-parallelism.
\subsection{Integration with Grid Operators}
Closer integration with grid operators could improve forecast accuracy, support participation in demand-response programs, and provide better visibility into when opportunistic migration windows are truly feasible.
\subsection{Extensions to Federated Learning}
The infrastructure naturally extends to federated learning, where migrating model updates rather than full checkpoints may place more workloads within the feasible region.
\subsection{Economic Models for Distributed AI}
Broad deployment requires economic mechanisms that align incentives across micro-datacenter operators, renewable producers, workload owners, network providers, and grid operators. Key components include: stakeholder-role analysis, valuation of renewable usage and avoided emissions, tokenized coordination primitives (compute credits, renewable-compute certificates, network-capacity incentives, reputation systems), and mechanisms for dynamic pricing and forecasting penalties. These baselines arise naturally from differing stakeholder objectives, underscoring the need for a unifying feasibility-aware framework. Any economic design must also interoperate with existing demand-response, renewable-energy certificate, and carbon-offset markets.
\parbreak
Overall, the evaluated baselines represent practical operational strategies; their limitations stem from structural constraints rather than flawed assumptions. Future work will formalize these economic dimensions and develop scalable coordination mechanisms that support large-scale, multi-stakeholder deployment.
\subsection{Threats to Validity and Limitations}
The feasibility-domain model abstracts away several aspects of real deployments. First, we assume stable effective WAN bandwidth between sites over the duration of a migration. In shared wide-area networks, background traffic and routing changes may temporarily reduce throughput and push some Class~A workloads towards the boundary of Class~B. Incorporating online bandwidth estimation and conservative margins partially mitigates this effect but does not eliminate it.
\parbreak
Second, our evaluation uses synthetic job classes derived from representative checkpoint sizes rather than a full production trace. While this allows controlled exploration of the feasibility space, it may under-represent corner cases such as highly bursty arrival patterns, mixed-precision training regimes, or workloads with atypical checkpointing strategies.
\parbreak
Third, we do not model failures during migration. In practice, node outages or network partitions may interrupt transfers, requiring rollback or re-scheduling. Extending the feasibility analysis to include reliability metrics and recovery costs is an important direction for future work, especially for deployments spanning multiple administrative domains.
}

\section{Conclusion}

This work introduced a quantitative feasibility-domain model for migratory AI workloads across renewable-powered micro-datacenters. We showed that energy breakeven is always reached within minutes, making \emph{time}—not energy—the dominant constraint. As a result, only workloads with small-to-moderate checkpoint sizes can be migrated within typical renewable windows at 1–10\,Gbps. Enforcing these boundaries enables the orchestrator to reduce non-renewable energy use by over 50\% while also improving JCT, whereas energy-only migration increases delay.

The approach is immediately applicable to the large class of single-GPU workloads whose checkpoints fall within the feasible region. Broader architectural and economic considerations are discussed in the extended version.

\section*{Acknowledgment}
   This work has received funding from the I-NEST EDIH under EU Grant Agreement n. 101083398).

\clearpage
\ifCLASSOPTIONcaptionsoff
  \newpage
\fi



\bibliographystyle{IEEEtran}
\bibliography{references}

\appendices
\section{Additional material}

\shortver{
\subsection{Boundary Conditions for Simulation}
The feasibility-domain validation in Section~\ref{sec:feasibility} relies on a realistic set of system and network parameters that reflect current micro-datacenter deployments and WAN interconnects. These boundary conditions define the operating envelope used in our 5-node simulation and determine both the migration timing and the achievable renewable-only execution windows.
Table~\ref{tab:boundary_conditions_short} reports the specific parameters used in the experiments, including WAN bandwidth, typical renewable surplus duration, and empirical values for checkpoint loading and downtime. These assumptions correspond to the feasibility limits established in Section~\ref{sec:feasibility}, ensuring that the simulation directly evaluates the orchestrator under realistic constraints.
\begin{table}[h]
\caption{Boundary Conditions for Example Scenario}
\label{tab:boundary_conditions_short}
\centering
\begin{tabular}{@{}llr@{}}
\toprule
\textbf{Parameter} & \textbf{Value} & \textbf{Source} \\ \midrule
WAN Capacity & 10 Gbps & AWS inter-region \\
Energy Surplus & 2.5h & CAISO avg. \\
Downtime & < 1s & \cite{Wei2025PhoenixOS} \\
Checkpoint Load & 10.3s & \cite{serverlessllm2024} \\
Acceptable Overhead & 10\% & Target \\
\bottomrule
\end{tabular}
\end{table}
A simple sensitivity check shows that increasing WAN capacity by $10\times$ shifts many Class~B workloads into Class~A, significantly expanding the feasible migration envelope.
}

\shortver{
\subsection{Workload Classification}
Based on the feasibility boundaries derived above—primarily determined by checkpoint size and WAN bandwidth—we group workloads into three practical classes. The classification reflects how quickly their training state can be migrated relative to typical renewable-energy windows and whether migration remains energetically profitable. Table~\ref{tab:workload_classes_short} summarizes the resulting categories, from highly suitable small-model workloads to large LLMs whose state size exceeds feasible migration limits.
\begin{table}[b]
\caption{Workload Classification by Migration Feasibility}
\label{tab:workload_classes_short}
\centering
\small
\begin{tabular}{@{}p{0.22\columnwidth}p{0.19\columnwidth}p{0.18\columnwidth}p{0.27\columnwidth}@{}}
\toprule
\textbf{Class} & \textbf{Type} & \textbf{Size} & \textbf{Feasibility} \\ \midrule
A: Suitable & Small & $<$10 GB & $<$1 min \\
B: Conditional & Medium & 10--100 GB & Minutes \\
C: Infeasible & Large LLMs & $>$100 GB & Exceeds window \\
\bottomrule
\end{tabular}
\end{table}
}

\subsection{Feasibility-Aware Migration Algorithm}
Algorithm~\ref{alg:scheduler_short} presents our feasibility-aware migration scheduler.
\begin{algorithm*}[!t]
\caption{Feasibility-Aware Migration Scheduler}
\label{alg:scheduler_short}
\begin{algorithmic}[1]
\For{each $\text{scheduling\_interval}$}
    \State $\text{energy\_forecast} \gets \text{GetRenewableForecasts()}$
    \State $\text{network\_state} \gets \text{MeasureInterSiteBandwidth()}$
    \For{each $\text{job}$ in $\text{running\_jobs}$}
        \State \textcolor{blue}{// --- FEASIBILITY CHECK ---}
        \State $T_{\text{transfer}} \gets \text{CalcTransferTime}(\text{job}, \text{network\_state})$
        \State $T_{\text{load}} \gets \text{GetLoadTime}(\text{job})$
        \State $T_{\text{cost\_time}} \gets T_{\text{transfer}} + T_{\text{load}} + 0.4\text{s}$
        \State
        \State $E_{\text{cost\_energy}} \gets \text{CalcEnergyCost}(\text{job})$
        \State $T_{\text{breakeven\_energy}} \gets E_{\text{cost\_energy}} / 0.75\text{ kW}$
        \State
        \State \textcolor{blue}{// Fail if either time constraint OR energy breakeven is violated}
        \If{$T_{\text{cost\_time}} > (0.1 \times \text{energy\_forecast.duration}) 
        \textbf{   OR   } 
        T_{\text{breakeven\_energy}} > \text{energy\_forecast.duration}$}
            \State \textbf{continue} \textcolor{blue}{// Not feasible}
        \EndIf
        \State
        \State \textcolor{blue}{// --- OPTIMIZATION (within feasible space) ---}
        \State $\text{benefit} \gets \text{CalcBenefit}(\text{job}, \text{sites})$
        \If{$\text{benefit} > T_{\text{cost\_time}}$}
            \State $\text{TriggerMigration}(\text{job}, \text{target\_site})$
        \EndIf
    \EndFor
\EndFor
\end{algorithmic}
\end{algorithm*}